\documentclass[a4paper,12pt,twoside]{article}
\usepackage{psfig}
\usepackage{epsfig}
\usepackage{amssymb,wasysym,bbm,feynmf,rotating,subfigure,here,citesort,bm}
\usepackage{graphicx}
\newcommand{\be}{\begin{equation}}
\newcommand{\ee}{\end{equation}}
\newcommand{\bea}{\begin{eqnarray}}
\newcommand{\eea}{\end{eqnarray}}
\newcommand{\benn}{\begin{displaymath}}
\newcommand{\eenn}{\end{displaymath}}
\newcommand{\beann}{\begin{eqnarray*}}
\newcommand{\eeann}{\end{eqnarray*}}
\newcommand\eref[1]{Eq.~(\ref{#1})}

\def\ga{{_{\mbox{\scriptsize G}}}}
\def\ng{{_{\mbox{\scriptsize NG}}}}
\def\nd{{_{\mbox{\scriptsize ND}}}}
\def\dd{{\mbox{\rm d}}}
\def\dD{{\mbox{\rm D}}}
\def\dP{\dd {\cal P}}
\def\DP{\dD {\cal P}}
\def\a{\alpha}
\def\d{\delta} 
\def\e{\epsilon} 
\def\g{\gamma} 
 
\def\z{\zeta} 
\def\bu{{\bm{u}}} 
\def\bv{{\bm{v}}} 
\def\bw{{\bm{w}}} 
\def\bx{{\bm{x}}}
\def\by{{\bm{y}}} 
\def\bz{{\bm{z}}} 
\newcommand{\gtsim}{\lower-0.45ex\hbox{$>$}\kern-0.77em\lower0.55ex\hbox{$\sim$}}
\newcommand{\ltsim}{\lower-0.45ex\hbox{$<$}\kern-0.77em\lower0.55ex\hbox{$\sim$}}
\newcommand{\befig}{\begin{figure}}
\newcommand{\efig}{\end{figure}}
\newcommand{\betab}{\begin{table}}
\newcommand{\etab}{\end{table}}

\def\del{\partial}                              

\begin{document}
%
%
%
\pagestyle{empty}
%
%
\title{ \vspace*{-1cm} {\normalsize\rightline{CU-TP-1125}}
  \vspace*{0.cm} 
{\Large \bf Extension of the JIMWLK equation in the low gluon
  density region}\footnote{This work is supported in part by the US
  Department of Energy.}}

\author{}
\date{} \maketitle
 
\vspace*{-2.5cm}
 
\begin{center}
 
\renewcommand{\thefootnote}{\alph{footnote}}
 
{\large
A.~H.~Mueller\,\footnote{arb@phys.columbia.edu (A.H.Mueller)},
A.~I.~Shoshi\,\footnote{shoshi@phys.columbia.edu} and 
S.~M.~H.~Wong\,\footnote{s\_wong@phys.columbia.edu}}
 
 
{\it Physics Department, Columbia University, New York, NY 10027, USA}

 
\end{center}
 

\begin{abstract}
  
  It has recently been realized that the Balitsky-JIMWLK equations
  have serious shortcomings as equations to be used in small-$x$
  evolution near the unitarity limit. A recent generalization of the
  Balitsky equations has been given which corrects these shortcomings.
  In this paper we present an equivalent discussion, but in terms of
  the JIMWLK equation where we show that a new (fourth order
  functional derivative) term should be included. We also present a
  stochastic version of the new equation which, however, has some
  unusual mathematical aspects which are not as yet well understood.

  \vspace{1.cm}
 
 
\medskip


\end{abstract}

%
%
%

\pagenumbering{roman}
\pagestyle{plain}
%
%
%
\pagenumbering{arabic}
\pagestyle{plain}
%
\makeatletter
\@addtoreset{equation}{section}
\makeatother
\renewcommand{\theequation}{\thesection.\arabic{equation}}

\newpage
\section{Introduction}
\label{Sec_Introduction}

The problem of high energy (small $x$) evolution near the unitarity
limit~\cite{Gribov:tu} is one of the most important and widely studied
problems in high energy QCD.  A good understanding of this problem is
necessary for understanding the energy dependence of the saturation
momentum and other properties of the color glass condensate
(CGC)~\cite{CGC}.  This in turn is important for good understanding of
such diverse quantities as the initial energy density produced just
after a high energy heavy ion collision and the $Q^2$ and
$x$-dependence of structure functions in deep inelastic scattering at
small $x$ and at moderate values of $Q^2.$

The main equations that have been used to deal with high energy
scattering near the unitarity limit are the Balitsky-JIMWLK
equations~\cite{Balitsky:1995ub+X,Jalilian-Marian:1997jx+X,Iancu:2001ad+X,Weigert:2000gi}
and the Kovchegov equation~\cite{Kovchegov:1999yj+X}. The Balitsky
equations are an infinite hierarchy of coupled equations expressing
the energy dependence of the scattering of high energy quarks and
gluons (represented by Wilson lines in the fundamental and adjoint
representations, respectively) on a target. The Kovchegov equation
results when the large-$N_c$ limit of the Balitsky hierarchy is taken
along with dropping all correlations in the target.  The Kovchegov
equation is a (nonlinear) equation for a single function and, while no
exact solution is known, the general properties of the solutions of
this equation near the unitarity boundary are
understood~\cite{Levin:1999mw+X,Armesto:2001fa+X,Golec-Biernat:2001if,Levin:2001yv+X,Golec-Biernat:2003ym,Albacete:2003iq,Albacete:2004gw,Iancu:2002tr,Mueller:2002zm,Triantafyllopoulos:2002nz,Munier:2003vc+X,Munier:2004xu,Bartels:2004jn,Chachamis:2004ab,Levin:2004yd}.
The JIMWLK equation is a functional Fokker-Planck equation which
expresses small-x evolution of the target in a way which is exactly
equivalent to the evolution of the scattering amplitudes given by the
Balitsky equations.  The JIMWLK equation is the basic equation for the
CGC.

Geometric scaling~\cite{Stasto:2000er} is a general property of
solutions to the Kovchegov equation, and the exact (asymptotic) energy
dependence of the saturation
momentum~\cite{Mueller:2002zm,Triantafyllopoulos:2002nz,Munier:2003vc+X}
is also known for the Kovchegov equation.  However, recently two of
the present authors argued that while guaranteeing that overall
scattering amplitudes satisfy unitarity limits the Kovchegov equation
allows significant unitarity violation in intermediate stages of
small-$x$ evolution~\cite{Mueller:2004se}.  These authors then
proposed a simple prescription for suppressing the unitarity
violations (a simple boundary limiting small size rare dipole
fluctuations) without, however, realizing the generality of their
prescription.  Later, in Ref.~\cite{Iancu:2004es}, it was realized
that the prescription was identical to a procedure used to study
reaction-diffusion systems in statistical
physics~\cite{BrunetDerrida,Saarlos}, and that the prescription has a
general validity for a certain class of observables such as the energy
dependence of the saturation momentum.  Again using the relationship
between small-$x$ evolution and the time dependence of
reaction-diffusion processes in statistical physics an asymptotic
scaling law for the scattering amplitude, strongly violating geometric
scaling was derived.  This scaling differs from, but is related to,
that found in Ref.~\cite{Mueller:2004se} where fluctuations were not
included in the purely mean field approach followed there.

\vskip 5pt Thus, the Kovchegov equation misses some essential aspects
of high energy evolution near the unitarity boundary, although it may
still be accurate for large nuclei or at intermediate energies.  In
this light the results of numerical
simulations~\cite{Rummukainen:2003ns} comparing evolution using the
Kovchegov equation with that using the Balitsky-JIMWLK equations come
as a big surprise, because no essential differences were found.  While
this is partly due to the initial (Gaussian) distribution used in
Ref.~\cite{Rummukainen:2003ns} the deeper understanding came from
Iancu and Triantafyllopoulos~\cite{Iancu:2004iy} who noted that the
Balitsky-JIMWLK equation also misses essential aspects of unitarity
constraints.  In the language of ``Pomerons'', Balitsky-JIMWLK has
Pomeron mergings but no Pomeron splittings.  The authors of
Ref.~\cite{Iancu:2004iy} then propose a generalization of the Balitsky
equations which includes a stochastic part, thus allowing for Pomeron
splittings, in addition to the usual Balitsky hierarchy.  (In the
context of Pomeron interactions this term, and its relationship to the
dipole model, has been known for some time~\cite{Bartels:1994jj+X}.)
It is likely that the Balitsky hierarchy augmented by this new term
contains the essential ingredients for small-$x$ evolution near the
saturation boundary.

The discussion of Ref.~\cite{Iancu:2004iy} is given completely in
terms of the Balitsky approach to high energy scattering. In this
paper we give the corresponding (physically equivalent) discussion in
the context of the JIMWLK approach.  That is, we give a new equation,
which includes the original contribution due to JIMWLK along with an
extra term, the resulting equation having the form of a functional
differential equation for the CGC weight function $W[\alpha]$. The
JIMWLK equation, given below in Eq.~(\ref{JIMWLK}), is a functional
Fokker-Planck equation which has up to two derivatives of $W[\alpha]$
with respect to $\alpha$ but, in addition, has multiplicative powers
of $\alpha$ such that there are always at least as many powers of
$\alpha$ multliplying $W[\alpha]$ as there are $\alpha$-derivatives
acting on $W[\alpha]$. The term having equal numbers of $\alpha$'s as
derivatives of $\alpha$ generates the kernel for BFKL evolution while
terms having extra powers of $\alpha$ correspond to ``Pomeron
mergings''.  The absence of terms having more derivatives than factors
of $\alpha$ is the absence of ``Pomeron splittings.''  The added new
term has precisely four derivatives of $\alpha$ along with two factors
of $\alpha$.  The exact nature of this term is determined by requiring
that it correspond to one parent dipole splitting into two daughter
dipoles in the dipole model.  If the parent dipole is part of a BFKL
evolution, and if the two daughter dipoles undergo BFKL evolution as
generated by the original terms in the JIMWLK equation, then our new
term can be viewed as a single Pomeron splitting into two Pomerons. In
Section 2 we carry out the construction of this splitting term with
the resulting modified JIMWLK equation given in Eq.~(\ref{JIMWLK_DM}).

Our construction of this new term is carried out in the
weak field limit where the dipole
model~\cite{Mueller:1994rr,Mueller:1994jq} is valid.  We have no
argument to the effect that it correctly represents Pomeron splittings
in a strong field environment.  On the other hand this new term is
only important, as compared with the original terms in the JIMWLK
equation, in the weak field limit so that we believe (\ref{JIMWLK_DM}) represents
the essential physics of BFKL evolution and unitarity constraints.
That is, we believe that (\ref{JIMWLK_DM}) is a good ``effective'' equation for
small-$x$ evolution.  If one wished to include higher order
BFKL-kernel effects in the JIMWLK equation one could expect
corrections to our new term also to be necessary, but in the context
of evolution using the lowest order BFKL kernel, (\ref{JIMWLK_DM}) should be
sufficiently accurate.  Iancu and Triantafyllopoulos~\cite{Iancu;Triant} have observed
that (\ref{JIMWLK_DM}) is exactly equivalent to their new equation, and we also
expect that it is equivalent to the procedure of unitarization
developed some time ago by Salam~\cite{Salam:1995zd}, at least in the energy regime where
that procedure is valid.

One of the very nice features of the JIMWLK equation is that it is
possible to view it as a Langevin equation and hence use it as a basis
for numerical simulations.  Indeed, successful simulations have been
performed by Rummukainen and Weigert~\cite{Rummukainen:2003ns}.  This
motivates trying to cast (\ref{JIMWLK_DM}) in the form of a stochastic
differential equation.  In Section 3 this is done with, however, a few
unusual aspects.  In general a fourth order derivative will require a
non-Gaussian noise, and this is illustrated by a simple example in
Section 3.1.  In Eq.~(\ref{eq:a-stoch-eqn}) we give the rule for the
evolution of $\alpha$ which, in a stochastic sense, is equivalent
to~(\ref{JIMWLK_DM}).  Eq.~(\ref{eq:a-stoch-eqn}) involves three noise
terms: $\zeta(\bm{z})$ is a fourth-order noise term,
$\bar{\nu}^a(\bm{x},\bm{\omega})$ is a Gaussian noise term and
$\xi(\bm{x},\bm{\omega})$ is a non-diagonal noise term whose
correlator is given in~(\ref{eq:nondiagonal-corr}).  While we have
been successful in casting our equation in the form of a stochastic
differential equation the unusual form of the noise terms makes it
unclear whether this form will be at all useful for numerical
simulation.  We do note, however, that the fourth-order noise terms
have been discussed in the mathematical
literature~\cite{Gardiner,Hochberg}, but we have not been able to find
discussions of terms like our non-diagonal noise, and we are not
certain that this term makes good mathematical sense.

\section{Weak field extension of the JIMWLK equation}
\label{Sec_E_JIMWLK}

Iancu and Triantafyllopoulos~\cite{Iancu:2004iy} have recently
recognized that the JIMWLK
equation~\cite{Jalilian-Marian:1997jx+X,Iancu:2001ad+X,Weigert:2000gi}
-- the basic equation of the Color Glass Condensate formalism --
includes Pomeron mergings but not also Pomeron splittings which are
important at low gluon densities (weak fields) in the wavefunction of
a target. In this Section we extend the JIMWLK equation in the weak
field regime by adding the Pomeron splittings.

\subsection{JIMWLK equation}
\label{Sec_JIMWLK}
The JIMWLK
equation~\cite{Jalilian-Marian:1997jx+X,Iancu:2001ad+X,Weigert:2000gi}
reads
\be
\frac{\partial}{\partial Y} W_Y[\alpha] = 
  \frac{1}{2} \int_{\bm{x}, \bm{y}}\,
  \frac{\delta}{\delta \alpha_Y^a(\bm{x})}\,
  \eta^{ab}(\bm{x},\bm{y})\,\frac{\delta}{\delta
    \alpha_Y^b(\bm{y})} W_Y[\alpha]\ ,
\label{JIMWLK}
\ee
where $Y = \ln 1/x$ is the rapidity of the small-$x$ gluons,
$\int_{\bm{x}} \equiv \int d^2 {\bm x}$ ($\bm{x}$, $\bm{y}$ denote
transverse coordinates), $\alpha^a$ is the color gauge field radiated
by the color sources in the target, and the functional $W_Y[\alpha]$
is the weight function for a given field configuration. The kernel
$\eta^{ab}({\bm{x}},{\bm{y}})$ is a non--linear functional of
$\alpha$,
\be
\eta^{a b}(\bm{x},\bm{y}) = 
  \frac{1}{\pi}\int_{\bm{z}}\,
  {\cal K}({\bm{x}, \bm{y}, \bm{z}})
  \,(1- \tilde V^\dagger_{\bm{z}}  \tilde V_{\bm{x}})^{fa}(1 -  \tilde V^\dagger_{\bm{z}}
  \tilde V_{\bm{y}})^{fb} \ ,
\label{eta}
\ee
as it depends on the Wilson lines in the adjoint representation $\tilde V$ and  $\tilde V^\dagger$,
\be
\tilde{V}^\dagger_{\bm{x}}[\alpha]
\,=\,{\rm P}\,{\rm exp}\left({\rm i}g\int dx^-
\alpha^a(x^-,{\bm{x}}) T^a\right) \ , 
\label{Vdef}
\ee
where $P$ denotes the path--ordering along the light-cone coordinate
$x^-$ and 
\be
{\cal K}({\bm{x}, \bm{y}, \bm{z}}) = \frac{1}{(2\pi)^2}\,
   \frac{(\bm{x}-\bm{z})\cdot(\bm{y}-\bm{z})}{
     (\bm{x}-\bm{z})^2 (\bm{z}-\bm{y})^2} \ .
\label{calK}
\ee

The average of a generic operator $\cal{O}$ is defined as 
\be
\langle {\cal O}\rangle_Y\,=\,
\int\,{\rm D}[\alpha]\, \,W_Y[\alpha]\,\,{\cal O}[\alpha] 
\label{OBS}   
\ee
and its evolution with respect to $Y$, 
%
\be
\frac{\del \langle {\cal O}\rangle_Y}{\del Y}\,=\,
\left\langle{1 \over 2}\int_{{\bm{x}},{\bm{y}}}\,
{\delta \over {\delta \alpha^a({\bm{x}})}}\,
\eta^{ab}({\bm{x}},{\bm{y}}) \,{\delta \over 
{\delta \alpha^b({\bm{y}})}}\,{\cal O}[\alpha]
\right\rangle_Y \ ,
\label{Oevol}
\ee
follows by first using (\ref{JIMWLK}) and then integrating twice by parts in (\ref{OBS}).

As an example, consider the scattering of a single dipole off a
target. In the eikonal approximation, the T-matrix is given by 
\be
T({\bm{x}},{\bm{y}})\,=\, 1- \frac{1}{N_c}\,
{\rm tr}\Big(V^\dagger({\bm{x}}) V({\bm{y}})\Big)\ ,
\label{Tdef}
\ee
where $V^\dagger({\bm{x}})$ ($V({\bm{y}})$) is a Wilson line in the
fundamental representation ($T^a \to t^a$ in (\ref{Vdef})) which
describes the scattering of the quark (antiquark) off the color field
in the target, and the color trace divided by $N_c$ is the average
over color. Inserting the above expression in Eq.~(\ref{Oevol}), one
obtains
\bea
\!\!\!\frac{\partial}{\partial Y} \langle T({\bm{x}},{\bm{y}}) \rangle_Y &=&
           \frac{\alpha_s N_c}{2 \pi^2} \int_{{\bm{z}}}
           \frac{(\bm{x}-\bm{y})^2}{
           (\bm{x}-\bm{z})^2 (\bm{z}-\bm{y})^2} \nonumber \\
            &&\Big\langle
           - T({\bm{x}},{\bm{y}}) + T({\bm{x}},{\bm{z}}) + 
            T({\bm{z}},{\bm{y}})-
            T({\bm{x}},{\bm{z}})T({\bm{z}},{\bm{y}}) \Big\rangle_Y \ .
\label{Eq_Balitsky}
\eea 
In this non-linear equation, the scattering of a single dipole
$\langle T({\bm{x}},{\bm{y}}) \rangle_Y$ on the l.h.s. is coupled to
the scattering of two dipoles off the target $\langle T^{(2)}\rangle
\equiv \langle T({\bm{x}},{\bm{z}})T({\bm{z}},{\bm{y}}) \rangle_Y$ on
the r.h.s., and, in general, the equation for $\langle T^{(N)}\rangle$
will involve also $\langle T^{(N+1)}\rangle$. So, Eq.~(\ref{JIMWLK})
generates an infinite hierarchy of coupled non-linear equations. The
same set of coupled equations has also been derived by
Balitsky~\cite{Balitsky:1995ub+X} in a frame where only the projectile
evolves with increasing rapidity. Taking the large $N_c$ limit of the
JIMWLK-Balitsky hierarchy (non-dipolar terms become neglegible) and
dropping the correlations in the target ($\langle
T({\bm{x}},{\bm{z}})T({\bm{z}},{\bm{y}})\rangle \to \langle
T({\bm{x}},{\bm{z}})\rangle\,\langle T({\bm{z}},{\bm{y}})\rangle$),
one obtains the Kovchegov equation~\cite{Kovchegov:1999yj+X} which is
a closed equation for the $T$-matrix and, thus, much easier to handle
than the coupled JIMWLK-Balitsky equations. Although no exact solution
is known, the general properties of the solution to the Kovchegov
equation are well
understood~\cite{Levin:1999mw+X,Armesto:2001fa+X,Golec-Biernat:2001if,Levin:2001yv+X,Golec-Biernat:2003ym,Albacete:2003iq,Albacete:2004gw,Iancu:2002tr,Mueller:2002zm,Triantafyllopoulos:2002nz,Munier:2003vc+X,Munier:2004xu}.

\subsection{JIMWLK equation in the weak field limit}
\label{Sec_Weak_Field_Approximation}
In the weak field limit ($\alpha \sim g\, \ltsim\, 1$), i.e., for low
gluon density in the target, one can expand the kernel $\eta^{ab}$
in Eq.~(\ref{eta}) in powers of $g$,
\be
\eta^{a b}({\bm{x}},{\bm{y}}) \,\simeq\,\frac{g^2\big(T^c T^d\big)_{ab}}{\pi}
\int_{\bm{z}}\, {\cal K}({\bm{x}, \bm{y}, \bm{z}})\,\big[
\alpha^c({\bm{x}}) - \alpha^c({\bm{z}})\big]\big[
\alpha^d({\bm{y}}) - \alpha^d({\bm{z}})\big] 
\label{etalin}
\ee
which when inserted in (\ref{Oevol}) yields 
\bea\label{Oevolwfa} 
\frac{\del \langle {\cal O}\rangle_Y}{\del Y} &=&
\frac{g^2\big(T^c T^d\big)_{ab}}{2\pi} \\
&\times& 
\!\!\!\!\int_{{\bm{x}},{\bm{y}},{\bm{z}}}\!\!\!\!\!{\cal K}({\bm{x}, \bm{y}, \bm{z}})\,
\!\left\langle
\!{\delta \over {\delta \alpha^a({\bm{x}})}}\,
\big[\alpha^c({\bm{x}})\!-\!\alpha^c({\bm{z}})\big]\big[
\alpha^d({\bm{y}})\!-\!\alpha^d({\bm{z}})\big] 
\,{\delta \over 
{\delta \alpha^b({\bm{y}})}}\,{\cal O}[\alpha]
\right\rangle_{\!Y} \ ,\nonumber
\eea 
where 
\be
\alpha^a({\bm{x}})\,\equiv\,\int dx^-\,\alpha^a(x^-,{\bm{x}}) \ .
\label{alphaT}
\ee
For the $T$-matrix (\ref{Tdef}) in the leading power in $g$,
\be
T_0({\bm{x}},{\bm{y}})\,= \frac{g^2}{4N_c}\, \Big(\alpha^a({\bm{x}})
- \alpha^a({\bm{y}})\Big)^2\,+\,{\cal O}(g^3),
\label{Texp_wf}
\ee 
equation (\ref{Oevolwfa}) leads to the dipole version of BFKL equation~\cite{Mueller:1994rr}
\be
\frac{\partial}{\partial Y} \langle T_0({\bm{x}},{\bm{y}}) \rangle_Y =
           \frac{\alpha_s N_c}{2 \pi^2} \int_{{\bm{z}}}
           \frac{(\bm{x}-\bm{y})^2}{
           (\bm{x}-\bm{z})^2 (\bm{z}-\bm{y})^2} 
           \Big\langle - T_0({\bm{x}},{\bm{y}}) + T_0({\bm{x}},{\bm{z}}) + 
            T_0({\bm{z}},{\bm{y}})\Big\rangle_Y \ .
\label{Eq_BFKL}
\ee
Note that in the weak field limit the JIMWLK equation (\ref{Oevolwfa})
conserves the number of fields: two $\alpha$'s are created by
$\alpha^2$ and two are destroyed by $(\delta / {\delta \alpha})^2$.
The BFKL equation (\ref{Eq_BFKL}) is an example which shows that the
evolution of a two-point correlation function (l.h.s) involves only
two-point correlation functions (r.h.s). More generally, in the weak
field limit, the evolution equation for an N-point correlation
function involves only N-point correlation functions. In the language
of perturbative QCD, the conservation of the number of fields
corresponds to processes in which no Pomeron splitting or Pomeron
merging in the t-channel is allowed.

\begin{figure}[h!]
\setlength{\unitlength}{1.cm}
\begin{center}
\epsfig{file=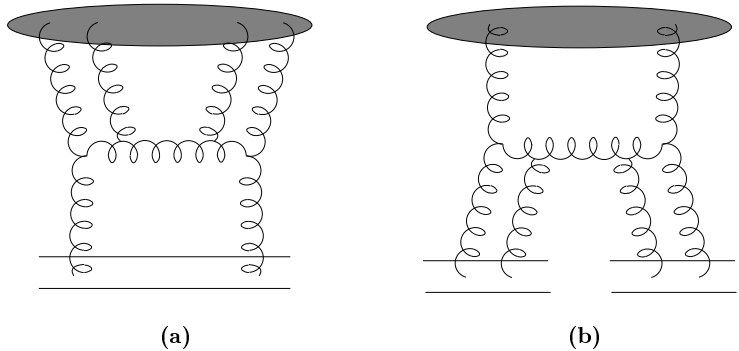, width=10cm}
\end{center}
\caption{Target evolution: (a) gluon merging, (b) gluon splitting.}
\label{Fig_Merging_Splitting}
\end{figure}

The JIMWLK equation describes gluon (Pomeron) mergings in the hadron
(nucleus) that are very important at high gluon densities (strong
fields). This happens through the non-linearities in the JIMWLK
equation: an N-point correlation function on the l.h.s. of
(\ref{Oevolwfa}) can involve in general all the M-point functions with
M $\ge$ N on the r.h.s. once the higher orders in $\alpha$ ($\alpha^n$ with
$n>2$) are taken into account in the expansion of $\eta^{ab}$ in
(\ref{etalin}). Gluon mergings reduce the number of fields in a
correlation function.  Fig.~\ref{Fig_Merging_Splitting}a shows the merging of $4$ gluons
into $2$ gluons (or a $4$-point function reduces to a $2$-point
function). A further BFKL evolution turns the $4$ to $2$ gluon merging
into a $2$ to $1$ Pomeron merging via the triple pomeron vertex.

It has recently been recognized~\cite{Iancu:2004iy} that the JIMWLK
equation does not provide gluon (Pomeron) splittings that turn out to
be important for low gluon densities or weak fields in the target.
This is because the JIMWLK equation (\ref{JIMWLK}) does not include
higher order derivatives $(\delta^n / {\delta \alpha^n})$ with $n>2$
which in the weak field limit (where $\eta^{ab} \propto
\alpha^2$) would generate gluon splittings. For instance, the JIMWLK
equation does not include $\alpha^2 (\delta^4 / {\delta \alpha^4})$
which would generate a splitting of $2$ gluons into $4$ gluons (a
$2$-point correlation function rises to a $4$-point correlation
function) as shown in Fig.~\ref{Fig_Merging_Splitting}b. Thus, for low
gluon densities in the target, there are no gluon number fluctations
(Pomeron splittings) in the JIMWLK equation.

\subsection{Extension of the JIMWLK equation in the weak field region based on the
  dipole model}
\label{Sec_JIMWLK_DM}
In the next steps we will extend the JIMWLK equation (\ref{JIMWLK}) in
the weak field regime by adding the gluon (Pomeron) splittings. We
include the gluon number fluctuations in the JIMWLK equation based on
the dipole model~\cite{Mueller:1994rr,Mueller:1994jq} which is
designed to properly describe the dilute region of a hadron
wavefunction.  The dipole model, on the other hand, has so far failed
to accommodate gluon mergings, important at high gluon densities, which
are a specialty of the JIMWLK equation.

To construct the weight function $W_Y[\alpha]$ of a target in the weak
field limit based on the dipole model, we first need to know
the weight function for a single dipole $W_{Y_0}[\alpha]$. The latter can
be obtained by requiring that it gives the well-known result for dipole-dipole
scattering in the $2$-gluon-exchange approximation when used to
evaluate the averaging for the $T_0$-matrix (\ref{Texp_wf}) according to
(\ref{OBS}). The following weight function for a single dipole with
coordinates ($\bm{x_0}, \bm{y_0}$)~\cite{Iancu:2003uh}
\be
W_{Y_0}[\alpha]\,=\,\left\{1 + \frac{g^2}{4N_c}
\int_{{\bm u},{\bm v}} \,{\cal G}({\bm u}|{\bm{x}}_0,{\bm{y}}_0)
{\cal G}({\bm v}|{\bm{x}}_0,{\bm{y}}_0)\,\frac{\delta^2}{\delta \alpha^a({\bm{u}})
\delta \alpha^a({\bm{v}})}\right\}\delta[\alpha] \ ,
\label{DWFexp}
\ee 
gives the desired result for dipole-dipole scattering,
\bea
\langle T_0({\bm{x}},{\bm{y}})\rangle_0 \,&=& 
\int\,{\rm D}[\alpha]\,
\,W_{Y_0}[\alpha]\,T_0({\bm{x}},{\bm{y}})\nonumber\\
&=& \frac{g^4 N_g}{8 N^2_c}\,
\Big({\cal G}({\bm x}|{\bm{x}}_0,{\bm{y}}_0) - {\cal G}({\bm
  y}|{\bm{x}}_0,{\bm{y}}_0)\Big)^2 \nonumber\\
&=&\frac{g^4 N_g}{8 N_c^2}\Big[\frac{1}{4 \pi}
\ln\frac{({\bm{x}}-{\bm{y}_0})^2({\bm{y}}-{\bm{x}_0})^2}
{({\bm{x}}-{\bm{x}_0})^2({\bm{y}}-{\bm{y}_0})^2}\Big]^2\ ,  
\label{Texp_wfr}
\eea
with the definitions 
\be
{\cal G}({\bm z}|{\bm{x}}_0,{\bm{y}}_0) = \Delta({\bm z}-{\bm{x}_0})
- \Delta({\bm z}-{\bm{y}_0}) \ , \quad \Delta(\bm{x}) = \int
\frac{d^2 \bm{k}}{(2\pi)^2}\, \frac{e^{i \bm{k}\cdot\bm{x}}}{\bm{k}^2} = \frac{1}{4 \pi}
\ln\frac{1}{{\bm{x}}^2 \mu^2} \ ,
\ee
and $N_g = N_c^2-1$. The infrared cutoff $\mu$ allows us to write down
the propagator but it is harmless as it cancels in the difference in
$\cal{G}$.

The weight function for N dipoles in the dipole model (DM) reads
\be
W^{\mbox{\tiny DM}}_Y[\alpha]=\sum_{N=1}^\infty\int\! d\Gamma_N\,\,P_N(Y)
\prod_{i=1}^N\!\bigg\{1+ \frac{\lambda}{2}\!
\int_{{\bm u},{\bm v}}\!\!{\cal G}_i({\bm{u}}){\cal G}_i({\bm{v}})
\,\frac{\delta^2}{\delta \alpha^a({\bm{u}})
\delta \alpha^a({\bm{v}})}\bigg\}\,\,\delta[\alpha]\ .
\label{WFYexp}
\ee
where 
\be
{\cal G}_i({\bm{u}}) \equiv {\cal G}({\bm{u}}|{\bm{x}_{i-1}},{\bm{x}_i})
\ee
with ${\bm{x}_{i-1}}$ and ${\bm{x}_i}$ the positions of the quark and
the antiquark in the $i$th dipole. The shorthand notation $P_N(Y)$
denotes the probability to find a system of N dipoles with the
transverse coordinates (${\bm{x}_{0}}$, ${\bm{x}_1}$), (${\bm{x}_{1}}$,
${\bm{x}_2}$), ...(${\bm{x}_{N-1}}$, ${\bm{y}_0}$) via the evolution of
an original dipole with coordinates (${\bm{x}_{0}}$, ${\bm{y}_0}$) up
to rapidity $Y$. The phase space integration is $d\Gamma_N =
d^2\bm{x}_1\,d^2\bm{x}_2 ...d^2\bm{x}_{N-1}$ and $\lambda = g^2/(2N_c)$.

The evolution of the weight function (\ref{WFYexp}) with increasing rapidity
$Y$ is given by the following linear evolution equation
\bea 
\frac{\partial}{\partial Y} W^{\mbox{\tiny DM}}_Y[\alpha]\!\!&=&\!\!
\frac{\alpha_s N_c}{2 \pi^2}\!\int_{\bm{x}, \bm{y}, \bm{z}}\frac{(\bm{x}-\bm{y})^2}
{(\bm{x}-\bm{z})^2 (\bm{z}-\bm{y})^2} \nonumber\\
&\times&\!\!\!\!\!
\Bigg[\,-
\Big(1 + \frac{\lambda}{2}
\int_{{\bm u},{\bm v}} \,{\cal G}({\bm u}|{\bm{x}},{\bm{y}})
{\cal G}({\bm v}|{\bm{x}},{\bm{y}})\,\frac{\delta^2}{\delta \alpha^a({\bm{u}})
\delta \alpha^a({\bm{v}})}\Big) \Big. \nonumber\\
&& +\Big(1 + \frac{\lambda}{2}
\int_{{\bm u_1},{\bm v_1}} \,{\cal G}({\bm u_1}|{\bm{x}},{\bm{z}})
{\cal G}({\bm v_1}|{\bm{x}},{\bm{z}})\,\frac{\delta^2}{\delta \alpha^a({\bm{u_1}})
\delta \alpha^a({\bm{v_1}})}\Big) \Big. \nonumber\\
&&\times
\Big(1 + \frac{\lambda}{2}
\int_{{\bm u_2},{\bm v_2}} \,{\cal G}({\bm u_2}|{\bm{z}},{\bm{y}})
{\cal G}({\bm v_2}|{\bm{z}},{\bm{y}})\,\frac{\delta^2}{\delta \alpha^b({\bm{u_2}})
\delta \alpha^b({\bm{v_2}})}\Big) \Bigg] \nonumber\\
&&\times \Big(\frac{1}{-2\lambda N_g}\Big)\, \nabla^2_{\bm{x}}\,\nabla^2_{\bm{y}} 
\,\alpha^c(\bm{x})\,\alpha^c(\bm{y})\, W^{\mbox{\tiny DM}}_Y[\alpha] \ .
\label{Eq_WDM}
\eea 
This equation is a sum of two terms: a virtual term (loss term) and a
real term (gain term) as we know them from the dipole model. The term
$\nabla^2_{\bm{x}}\,\nabla^2_{\bm{y}}\,\alpha(\bm{x})\,\alpha(\bm{x})$
plays the role of an annihilation operator for a dipole while
$(1+\lambda/2 \int {\cal G}^2 (\partial/\partial \alpha)^2)$
corresponds to a creation operator for a dipole. In the virtual term
a dipole is destroyed and its coordinates are transfered to the
new one which is created. In the real term a dipole is converted into two new
ones with the right coordinates and splitting factors.

To explicitly show the annihilation and creation of dipoles, consider the
evaluation of (\ref{Eq_WDM}) with an arbitrary function ${\cal
  O}[\alpha]$
\be
{\cal I} \equiv 
\int\,{\rm D}[\alpha]\, \, \left(\frac{\del W^{\mbox{\tiny
      DM}}_Y[\alpha]}{\del Y}\right)\,\, {\cal O}[\alpha]\ .
\label{Oevol_DM}
\ee
$\cal{I}$ will be zero unless one takes the functional derivatives
from one of the dipoles in (\ref{WFYexp}) and integrates them by parts
in (\ref{Oevol_DM}) on the $\alpha(\bm{x})\,\alpha(\bm{y})$ term.
Using 
\bea
&& \!\!\!\!\frac{\lambda}{2}\,
\int_{{\bm u},{\bm v}}\!{\cal G}_i({\bm{u}}){\cal G}_i({\bm{v}})
\,\frac{\delta^2}{\delta \alpha^a({\bm{u}})\delta \alpha^a({\bm{v}})} 
\,\nabla^2_{\bm{x}}\,\nabla^2_{\bm{y}} 
\,\alpha^c(\bm{x})\,\alpha^c(\bm{y})
\nonumber \\
&&\hspace{1cm}= \lambda N_g 
\left[\delta(\bm{x}-\bm{x_{i-1}})-\delta(\bm{x}-\bm{x_{i}})\right]
\left[\delta(\bm{y}-\bm{x_{i-1}})-\delta(\bm{y}-\bm{x_{i}})\right] \nonumber
\eea
one gets
\bea 
\frac{\partial}{\partial Y} W^{\mbox{\tiny DM}}_Y[\alpha]\!\!&=&\!\!
\frac{\alpha_s N_c}{2 \pi^2}\,\sum_{i=1}^N\,
\int_{\bm{x_{i-1}}, \bm{x_i}, \bm{z}}\frac{(\bm{x_i}-\bm{x_{i-1}})^2}
{(\bm{x_{i-1}}-\bm{z})^2 (\bm{z}-\bm{x_i})^2} \label{Eq_WDM_Ev}
\\
&\times&\!\!\!\!\! 
\Bigg[\,-
\Big(1 + \frac{\lambda}{2}
\int_{{\bm u},{\bm v}} \,{\cal G}({\bm u}|{\bm{x_{i-1}}},{\bm{x_i}})
{\cal G}({\bm v}|{\bm{x_{i-1}}},{\bm{x_i}})\,\frac{\delta^2}{\delta \alpha^a({\bm{u}})
\delta \alpha^a({\bm{v}})}\Big) \Bigg. \nonumber\\
&& +\Big(1 + \frac{\lambda}{2}
\int_{{\bm u_1},{\bm v_1}} \,{\cal G}({\bm u_1}|{\bm{x_{i-1}}},{\bm{z}})
{\cal G}({\bm v_1}|{\bm{x_{i-1}}},{\bm{z}})\,\frac{\delta^2}{\delta \alpha^a({\bm{u_1}})
\delta \alpha^a({\bm{v_1}})}\Big) \Big. \nonumber\\
&&\times
\Big(1 + \frac{\lambda}{2}
\int_{{\bm u_2},{\bm v_2}} \,{\cal G}({\bm u_2}|{\bm{z}},{\bm{x_i}})
{\cal G}({\bm v_2}|{\bm{z}},{\bm{x_i}})\,\frac{\delta^2}{\delta \alpha^b({\bm{u_2}})
\delta \alpha^b({\bm{v_2}})}\Big) \Bigg] \nonumber\\
&&\times \sum_{N=1}^\infty\int\! d\Gamma_N^{\prime}\,\,P_N(Y)
\prod_{{j=1}\atop {j\neq i}}^N\!\bigg\{1+ \frac{\lambda}{2}\!
\int_{{\bm u},{\bm v}}\!\!{\cal G}_j({\bm{u}}){\cal G}_j({\bm{v}})
\,\frac{\delta^2}{\delta \alpha^c({\bm{u}})
\delta \alpha^c({\bm{v}})}\bigg\}\,\,\delta[\alpha] , \nonumber
\eea 
where $d\Gamma_N^{\prime}$ means that the $\bm{x_{i-1}}$ and
$\bm{x_i}$ coordinate integrations are missing. By comparing this
equation with (\ref{Eq_WDM}), we now explicitly see that the evolution
is the same one known from the dipole model: When increasing the rapidity
in one step, the $N$-dipole system either survives as it is (virtual
term) or it evolves by splitting one of the dipoles into two new
dipoles (real term). In the latter, the kernel (times $\alpha_s
N_c/(2\pi^2)$) in the first line in (\ref{Eq_WDM_Ev}) is the right
factor to turn $P_N$ into $P_{N+1}$ while the product of the terms in
the third and fourth line are what is needed for converting the
original $\bm{i}$th dipole into two new dipoles ($\bm{x_{i-1}}$,
$\bm{z}$) and ($\bm{z}$, $\bm{x_i}$).

One can rewrite the evolution equation (\ref{Eq_WDM_Ev}) also in the form 
\be
\frac{\del W^{\mbox{\tiny DM}}_Y[\alpha]}{\del
  Y}=\sum_{N=1}^\infty\int\! d\Gamma_N\,\,\frac{\del P_N(Y)}{\del Y}
\prod_{i=1}^N\!\bigg\{1+ \frac{\lambda}{2}\!
\int_{{\bm u},{\bm v}}\!\!{\cal G}_i({\bm{u}}){\cal G}_i({\bm{v}})
\,\frac{\delta^2}{\delta \alpha^a({\bm{u}})
\delta \alpha^a({\bm{v}})}\bigg\}\,\,\delta[\alpha] \ ,\\
\label{DWDP}
\ee
which is consistent with Eq.~(\ref{WFYexp}). The identification of
(\ref{DWDP}) with (\ref{Eq_WDM_Ev}) gives the evolution equation for the
probabilities (see~\cite{Iancu:2003uh})   
\bea
\frac{\del P_N(Y|\bm{x_1},...,\bm{x_{N-1}})}{\del Y} =\!\!\! 
&-&\!\!\! \frac{\alpha_s N_c}{2\pi^2} \bigg[\sum_{i=1}^N 
\int_{\bm z} \frac{(\bm{x}_{i-1}-\bm{x}_i)^2}{(\bm{x}_{i-1}-\bm{z})^2 (\bm{x}_i-\bm{z})^2}\bigg]
P_N(Y|\bm{x_1},...,\bm{x_{N-1}})\nonumber \\
&&\hspace*{-5.cm}+\,\frac{\alpha_s N_c}{2\pi^2}\sum_{i=1}^{N-1} 
\frac{(\bm{x}_{i-1}-\bm{x}_{i+1})^2}{(\bm{x}_{i-1}-\bm{{x}_i})^2 (\bm{x}_{i+1}-\bm{{x}_i})^2}
\,\,P_{N-1}(Y|\bm{x_1},...,\bm{x_{i-1}},\bm{x_{i+1}}...,\bm{x_{N-1}}) \ , 
\label{evolP}
\eea
where $P_N(Y|\bm{x_1},...,\bm{x_{N-1}})$ denotes the probability for
$N$-dipoles with coordinates (${\bm{x}_{0}}$, ${\bm{x}_1}$),
(${\bm{x}_{1}}$, ${\bm{x}_2}$), ...(${\bm{x}_{N-1}}$, ${\bm{y}_0}$)
generated by the evolution of an original dipole with coordinates
($\bm{x_0}$, $\bm{y_0}$). The first term is a loss term which
describes the emission of one gluon from the original $N$-dipole
system while the second term is a gain term which describes the
formation of the $N$-dipole system via the splitting of one dipole in
an original $(N-1)$-dipole system. It is the loss term which takes
into account gluon (Pomeron) splittings missed in the JIMWLK
equation~\cite{Iancu:2004iy,Salam:1995zd}. Furthermore,
it is easy to show that the probability is conserved by
(\ref{evolP})~\cite{Iancu:2003uh}: $\sum_{i=1}^N \int d{\Gamma}_N P_N(Y) = 1$.

Let's now compare the evolution equation (\ref{Eq_WDM}) which follows
from the dipole model with the JIMWLK equation in the weak field
limit. The three $\cal{G}\cal{G}$ terms in (\ref{Eq_WDM})
correspond to ordinary BFKL evolution while the
$\cal{G}\cal{G}\cal{G}\cal{G}$ term allows Pomeron splitting.  Since
the ordinary BFKL terms were already properly included in the JIMWLK
equation, only the final, $\cal{G}\cal{G}\cal{G}\cal{G}$, term has to
be added to the JIMWLK equation. Thus, the main result of this work
reads
\bea
&&\!\!\!\!\!\!\!\!\frac{\partial}{\partial Y} W_Y[\alpha] = 
 \Bigg(\frac{1}{2} \int_{\bm{x}, \bm{y}}\,
  \frac{\delta}{\delta \alpha_Y^a(\bm{x})}\,
  \eta^{ab}(\bm{x},\bm{y})\,\frac{\delta}{\delta 
    \alpha_Y^b(\bm{y})} \nonumber \\ 
&&\!\!\!\!\!\!\!-\,\frac{N_c^2}{8\pi^3 N_g}\!\int_{\textstyle
  {\hspace{-0.9cm}{\bm{x},\bm{y},\bm{z}} 
\atop {\bm{u_1},\bm{v_1},\bm{u_2},\bm{v_2}}}} 
\!\!\!\!\!\left(\frac{\lambda}{2}\right)^2\!\!\!\frac{(\bm{x}-\bm{y})^2}
{(\bm{x}-\bm{z})^2 (\bm{z}-\bm{y})^2} 
\,{\cal G}({\bm u_1}|{\bm{x}},{\bm{z}})
{\cal G}({\bm v_1}|{\bm{x}},{\bm{z}})
{\cal G}({\bm u_2}|{\bm{z}},{\bm{y}})
{\cal G}({\bm v_2}|{\bm{z}},{\bm{y}}) \nonumber \\
&&\!\!\!\!\!\!\times
\,\frac{\delta}{\delta \alpha^a({\bm{u_1}})}
\,\frac{\delta}{\delta \alpha^a({\bm{v_1}})}
\,\frac{\delta}{\delta \alpha^b({\bm{u_2}})}
\,\frac{\delta}{\delta \alpha^b({\bm{v_2}})} 
\,\nabla^2_{\bm{x}}\,\nabla^2_{\bm{y}}
\,\alpha^c(\bm{x})\,\alpha^c(\bm{y})\Bigg)\,
W_Y[\alpha]\ . 
\label{JIMWLK_DM}
\eea
Our derivation of the Pomeron splitting term,
$\cal{G}\cal{G}\cal{G}\cal{G}$ term in (\ref{JIMWLK_DM}), is done in
the weak field limit. Therefore we cannot say that it also correctly
describes Pomeron splittings in the presence of strong fields
($\alpha \sim 1/g$), or high gluon densities, in the target.  However,
since the Pomeron splitting term is important only in the weak field
limit ($\alpha \sim g$) as compared with the original JIMWLK term, we
believe that (\ref{JIMWLK_DM}) is a good ``effective'' equation for
small-$x$ evolution which includes the essential physics of BFKL
evolution and unitarity constraints. Our Pomeron splitting term should
be accurate enough as long as the evolution involves only the lowest
order BFKL-kernel. Corrections to the new term are expected if higher
order BFKL-kernel effects are included in the original JIMWLK term.

The extended JIMWLK equation~(\ref{JIMWLK_DM}) includes Pomeron
mergings and Pomeron splittings and, thus, through iterations, Pomeron
loops. This equation is equivalent to the new equation derived by
Iancu and Triantafyllopoulos~\cite{Iancu;Triant}, and it should also
be equivalent to the procedure of unitarization developed by
Salam~\cite{Salam:1995zd} in which the gluon number fluctuations are
properly treated. We expect that the solution to Eq.~(\ref{JIMWLK_DM})
will confirm the results of the recent
works~\cite{Mueller:2004se,Iancu:2004es}. In these works it has been
shown that due to the fluctuations the rate of growth of the
saturation momentum (the ``saturation exponent'') considerably
decreases and the geometric scaling~\cite{Stasto:2000er} property
is strongly violated as compared to the numerical results of the
Balitsky-JIMWLK equation and the Kovchegov equation given
in~\cite{Rummukainen:2003ns}. The latter equations miss the
fluctuations or the Pomeron loops.

Below we write the extended JIMWLK equation~(\ref{JIMWLK_DM}) in a
form which is more appropriate for its stochastic interpretation
given in the next Section~\cite{Weigert:2000gi,Blaizot:2002xy}:
\bea
&&\!\!\!\!\!\!\!\!\frac{\partial}{\partial Y} W_Y[\alpha] = 
 \Bigg(-\frac{\delta}{\delta \a^a(\bx)}\, \sigma^a(\bx) + \frac{1}{2} \int_{\bm{x}, \bm{y}}\,
  \frac{\delta^2}{\delta \alpha_Y^a(\bm{x})\delta \alpha_Y^b(\bm{y})}\,
  \eta^{ab}(\bm{x},\bm{y}) \nonumber \\ 
&&\!\!\!\!\!\!\!-\,\frac{N_c^2}{8\pi^3 N_g}\!\int_{\textstyle
  {\hspace{-0.9cm}{\bm{x},\bm{y},\bm{z}} 
\atop {\bm{u_1},\bm{v_1},\bm{u_2},\bm{v_2}}}} 
\!\!\!\!\!\left(\frac{\lambda}{2}\right)^2\!\!\!\frac{(\bm{x}-\bm{y})^2}
{(\bm{x}-\bm{z})^2 (\bm{z}-\bm{y})^2} 
\,{\cal G}({\bm u_1}|{\bm{x}},{\bm{z}})
{\cal G}({\bm v_1}|{\bm{x}},{\bm{z}})
{\cal G}({\bm u_2}|{\bm{z}},{\bm{y}})
{\cal G}({\bm v_2}|{\bm{z}},{\bm{y}}) \nonumber \\
&&\!\!\!\!\!\!\times
\,\frac{\delta}{\delta \alpha^a({\bm{u_1}})}
\,\frac{\delta}{\delta \alpha^a({\bm{v_1}})}
\,\frac{\delta}{\delta \alpha^b({\bm{u_2}})}
\,\frac{\delta}{\delta \alpha^b({\bm{v_2}})} 
\,\nabla^2_{\bm{x}}\,\nabla^2_{\bm{y}}
\,\alpha^c(\bm{x})\,\alpha^c(\bm{y})\Bigg)\,
W_Y[\alpha]\ ,
\label{JIMWLK_DM_FP}
\eea
with 
\be 
\sigma^a(\bx) = \frac{1}{2} \int_\by \frac{\delta}{\delta \a^b(\by)} 
                    \eta^{ab}(\bx,\by) \ .
\label{sigma_pot}
\ee
The original JIMWLK equation (first two terms in (\ref{JIMWLK_DM_FP}))
is the familiar functional Fokker-Planck
equation~\cite{Weigert:2000gi,Blaizot:2002xy} which describes a
diffusion process with $\sigma^a(\bx)$ giving the drift term and
$\eta^{a,b}(\bx, \by)$ the diffusion coefficient. A Langevin equation
for the fields $\a(x)$ can be written down which is equivalent to the
original JIMWLK equation~\cite{Blaizot:2002xy}. The extended JIMLWK
equation (\ref{JIMWLK_DM_FP}) is more complicated as it contains also
the fourth order derivatives with respect to $\a$ (Pomeron splitting)
in addition to the familiar Fokker-Planck equation. In the next
Section we show how to obtain a stochastic differential equation also for the
extended JIMWLK equation. The hope hereby is that one may use the
stochastic differential equation for numerical simulations of
(\ref{JIMWLK_DM_FP}) as done for the original JIMWLK equation
in~\cite{Rummukainen:2003ns}.

\section{Interpretation of the evolution equation in terms of noise}

A functional equation like the JIMWLK equation although written in an
elegant compact form (not so compact after adding our extension for a
dilute color medium) is hard to use in actual computations. But this
obstacle can be surmounted via the observation that JIMWLK resembles
very much a diffusion equation~\cite{Blaizot:2002xy} which describes
the time evolution of molecules undergoing Brownian motion. The latter
has the interpretation of particles subjected to random acceleration
in the medium, in other words a Langevin interpretation of the motion
of the particles. In the case of JIMWLK, it is the light-cone gauge
fields $\a^a(\bx)$ that acquire a Gaussian random noise. In the
previous section it was shown that in the case of a weak field a
fourth order differential term was required to fully take into account
all the important contributions at the leading log level.
Unfortunately the extended JIMWLK equation ceases to be a functional
diffusion equation of the Fokker-Planck type and the Langevin
description is ruined.  In this section we aim to restore the
stochastic description even though the Langevin interpretation is
mostly irreversibly lost. This requires a multi-random noise
description involving both Gaussian white noise as well as other
higher order noise. We will first show the basics of how a purely
fourth order differential equation can arise out of a non-Gaussian
random noise introduced at the fundamental level of the equation of
motion.  Then progressing in increasing complexity a Langevin term
will be added which is well known to be responsible for the second
order diffusion equation. Thus one will eventually have a multi-noise
stochastic differential equation as the equation of motion from which
the extended JIMWLK equation can be recovered.

\subsection{Illustration with some simple toy models}

To demonstrate the essentials of how to arrive at a fourth order 
differential equation using random noise, we consider a system 
of particles whose motions, similar to molecules undergoing Brownian 
motion, are Markov in nature. That means what happens to the system of 
particle next in time depends only on the present state {\em and not} 
in any way on the history of this system of particles. As we will see,
this crucial condition simplifies the probabilistic description
tremendously when the probability density distribution at one
time can be related to a previous time in a relatively simple manner. 
Thus it is tempting to try to stay within the framework of 
Brownian motion, however we shall see that this is not really 
possible.

\subsubsection{A 1+1 dimensional system} 

To start off, let's consider the probability distribution, $P(x,t)$, of 
our system of particles which is a function of position $x$ and time 
$t$. The motion of the particles is taken to obey the stochastic
equation 
\be \dot x (t) = \sigma(x) + (\g(x))^{1/4}\, \z(t)  
\label{eq:1+1-stoch-eqn} 
\ee
where $\sigma(x)$ is the deterministic mean drift of the particles, 
$\g(x)$ is some function of $x$, and $\z(t)$ is a {\em non-Gaussian} 
random noise which has a normalized distribution (suppressing for 
the moment the $t$ dependence) 
\be \dP_\ng(\z) = f_\ng(\z)\, \dd \z  \;, 
\ee
so that 
\be \int f_\ng(\z)\, \dd\z  = 1 \;. 
\label{eq:4th-dist-norm} 
\ee 
Using the notation 
\be  \langle {\cal O}(\zeta) \rangle_\ng 
   = \int {\cal O}(\zeta) f_\ng (\zeta)\, \dd \zeta  
\ee
to denote an average over the distribution $f_\ng(\z)$, 
the random noise obeys 
\be \langle \z \rangle_\ng = \langle \z^2 \rangle_\ng 
   =\langle \z^3 \rangle_\ng = 0 \;, 
\label{eq:4th-corr-vanish}
\ee
and is only nonzero first in the four $\z$ correlator 
\be \langle \z^4 \rangle_\ng = c_\ng  
\label{eq:4th-corr-nonvanish}
\ee 
where $c_\ng$ is a number. This is the reason that $\z$ is referred 
to as a fourth order noise in the literature~\cite{Gardiner,Hochberg}. 
In that sense the usual Gaussian noise is second order noise. 
Note that the distribution $f_\ng(\z)$ cannot be written 
down simply as a function of some compact form but must be 
expressed in terms of a Fourier transform. The explicit expression 
$f_\ng(\z)$ serves little purpose in our discussion here and will not 
be written down. What is important here are the correlators given above, 
the first nonvanishing correlator is the one with four $\z$'s 
whereas it is the two noise correlator in the case of the Gaussian 
noise that is the first nonvanishing one. 

Given that the particle motion is Markov in nature, the probability
distribution of the particles $P(x,t)$ must depend on $P(x',t')$
where the particles currently at $x$ arrived there by previously first
reaching $x'$ at an earlier time $t'$. Bayes' theorem states the 
conditional probability $P(A|B)$ relating the probability $P(A)$ of an 
event A happening given that another related event B has already happened 
with probability $P(B)$ must satisfy  
\be P(A) = P(A|B) P(B) \;. 
\ee 
It follows that
\be P(x,t) = \int \dd x' P(x,t|x',t') P(x',t')  \;. 
\label{eq:1+1-cond-prob}
\ee
For a small time increment $\d t$ so that $t'=t-\d t$, the particle
motion, from \eref{eq:1+1-stoch-eqn}, changes by 
\be \d x(t) = x-x' = \Big [\sigma(x')+(\g(x'))^{1/4}\,\z(t) \Big ]\,\d t \;. 
\ee
Since $\z$ is a stochastic variable given the present position $x$, 
the previous $x'$ is therefore non-deterministic but has a distribution. 
The conditional probability $P(x,t|x',t')$ is therefore given by 
\be P(x,t|x',t') = \int \dP_\ng (\z(t)) \, 
    \d (x-x'-[\sigma(x')+(\g(x'))^{1/4}\,\z(t) ]\,\d t ) \;. 
\ee
In the limit that $\d t$ becomes very small, one can Taylor expand 
the $\d$-function in powers of $\d t$, but one has to keep terms up
to the fourth power of $\d t$ if they are accompanied by a 
corresponding power of $\z$ 
\bea
    P(x,t|x',t') \simeq \int \dP_\ng (\z(t)) 
            \!\!\!\!& &\!\!\!\!
    \Big \{ 
    \d(x-x')-\d'(x-x') \Big [\sigma(x')+(\g(x'))^{1/4}\,\z(t) \Big ]  \,\d t 
          \nonumber \\ & & 
   +\frac{1}{2} \d^{''}(x-x')
                       \Big [\sigma(x')+(\g(x'))^{1/4}\,\z(t) \Big ]^2\,\d t^2 
          \nonumber \\ & & 
   -\frac{1}{3!}\d^{'''}(x-x')
                       \Big [\sigma(x')+(\g(x'))^{1/4}\,\z(t) \Big ]^3\,\d t^3
          \nonumber \\ & & 
   +\frac{1}{4!}\d^{''''}(x-x')
                       \Big [\sigma(x')+(\g(x'))^{1/4}\,\z(t) \Big ]^4\,\d t^4 
          \nonumber \\ & & 
   + \dots \Big \}  \;.
\eea 
Here we have written 
$\del \d(x-x')/\del x= \d'(x-x') = -\del \d(x-x')/\del x'$. In view of 
\eref{eq:4th-corr-vanish} and \eref{eq:4th-corr-nonvanish}, the above
expansion can be simplified to 
\bea 
\!P(x,t|x',t') \simeq \d(x-x')-\d'(x-x')\, \sigma(x')\, \d t 
           + \frac{1}{4!} c_\ng \d^{''''}(x-x')\, \g(x')\, \d t^4  
           + \dots \;. 
\label{eq:1+1-expand} 
\eea 
Substituting this into \eref{eq:1+1-cond-prob} and expanding 
\be P(x',t-\d t) \simeq P(x',t) -\frac{\del P(x',t)}{\del t} \d t
                        +\dots  
\ee
as well, after some rearrangements and using 
$c_\ng = -1/\d t^3$, one finally has 
\be \frac{\del P(x,t)}{\del t} 
   = -\frac{\del\, \sigma(x) P(x,t)}{\del x} 
     -\frac{1}{4!} \frac{\del^4\, \g(x) P(x,t)}{\del x^4} \;. 
\ee 
The negative sign of $c_\ng$\footnote{This $c_\ng$ is given by the 
inverse of the third power of the small timestep $\d t$ is by no means 
accidental. From \eref{eq:4th-corr-nonvanish} one can deduce that 
$\z \propto \d t^{-3/4}$ or $\dd V =\z \dd t \propto \dd t^{1/4}$. 
The last would be a more or less standard expression for the non-Gaussian 
stochastic variable $V$ in the language of stochastic calculus. Equally
one can arrive at \eref{eq:1+1-expand} by applying Ito's lemma 
for fourth order noise. As one can see the equation has been 
expanded up to order $\d t^4$ only in appearance. In reality 
one has only reached the linear order in $\d t$.} is necessary for 
the stable evolution of $P(x,t)$. One can see that fourth order 
diffusion can be associated with a fourth order noise much in the 
same way that the Gaussian white noise gives rise to the second order 
diffusion equation.

\subsubsection{A 2+1 dimensional system} 

Let's now consider a system that has an evolution equation for its  
probability distribution that bears more resemblance to the equation
in Sec 2. Since the CGC evolution equation exists both in the functional 
space which is really infinite dimensional as well as the color space 
of QCD, we extend our system to two spatial dimensions. This should
suffice to demonstrate the main features of the noise interpretation
of our equation. In addition JIMWLK is basically a diffusion equation so 
there must be a stochastic term for the white noise as well. We will 
let the motion of the particles in this system obey 
\be
   \dot x^i (t) = \sigma^i(\bx) + \e^{ij}(\bx)\, \nu^j(t) 
                 +(\g(\bx))^{1/4}\,G_{ab}(\bx) \bar \nu^i_a(t)\,\z_b(t) \;. 
\label{eq:2+1-stoch-eqn}
\ee
$\sigma^i(x)$ is the deterministic drift as before, which is followed
by the term of the Gaussian white noise $\nu^j$ with the tensor 
coefficient function $\e^{ij}(\bx)$. $\nu^i$ has the usual normalized 
Gaussian distribution
\be d{\cal P}_\ga (\nu) = f_\ga(\nu) \, \dd^2\nu \;, 
\ee
\be \int f_\ga (\nu)\, \dd^2\nu  = 1 \;. 
\label{eq:dist-norm} 
\ee 
and expectation 
\be  \langle {\cal O}(\nu) \rangle_\ga 
   = \int {\cal O}(\nu) f_\ga (\nu)\, \dd^2\nu \;. 
\ee
The white noise $\nu^i(t)$ of course has all the usual properties 
\be \langle \nu^i \rangle_\ga = 0 
\label{eq:2nd-corr-vanish}
\ee
and 
\be \langle \nu^i \nu^j \rangle_\ga = c_\ga\, \d^{ij} \;. 
\label{eq:2nd-corr-nonvanish}
\ee
The last term of \eref{eq:2+1-stoch-eqn} is the fourth order 
noise term but now with more complicated functions as its coefficient  
as well as a set of secondary random Gaussian distributed variables  
$\bar \nu^i_a$. The last have the normalized distribution  
\be \dd \bar P_\ga (\bar \nu_a) = \bar f_\ga(\bar \nu_a) \,
                                  \dd^2\bar \nu_a \;, 
\ee
which gives 
\be \langle \bar \nu^i_a \bar \nu^j_b \rangle_\ga = \d^{ij}\, \d_{ab} \;. 
\label{eq:2nd-bar-corr-nonvanish} 
\ee

The form of \eref{eq:2+1-stoch-eqn} is peculiar but is necessary
to arrive at the extended evolution equation. Proceeding in a similar
fashion to the previous subsection with a small time increment 
\be  \d x^i (t) = x^i-x'^{i} 
                = \Big [ \sigma^i(\bx') + \e^{ij}(\bx')\, \nu^j(t) 
                        +(\g(\bx'))^{1/4}\,G_{ab}(\bx') \bar \nu^i_a(t)\,
                         \z_b(t)
                  \Big ] \d t \;. 
\ee
Bayes' theorem gives us 
\be P(\bx,t) = \int \dd^2 \bx' P(\bx,t|\bx't') P(\bx',t')  \;, 
\label{eq:2+1-cond-prob}
\ee
with the conditional probability 
\bea P(\bx,t|\bx't') \!&=&\! \int \dP_\ga (\nu)\,
                        \prod_a \dd \bar {\cal P}_\ga (\bar \nu_a)\,
                        \dP_\ng (\z)              \nonumber \\    
                              & &\!\! \times 
    \d^2 (\bx-\bx'-[ \sigma^i(\bx') + \e^{ij}(\bx')\, \nu^j(t) 
                    +(\g(\bx'))^{1/4}\,G_{ab}(\bx') \bar \nu^i_a(t)\,
                     \z_b(t)
               ] \d t )                                   \nonumber \\   
\eea
which once again can be expanded or one can equally apply Ito's lemma. 
Bearing in mind all the vanishing correlators of the different noise, 
we have  
\bea P(\bx,t|\bx't') = \!\!\!\!& &\!\!\!\!
                       \int \dP_\ga (\nu)\,
                       \prod_a \dd \bar {\cal P}_\ga (\bar \nu_a)\,
                       \dP_\ng (\z)              
      \nonumber \\  & &  
    \Big \{ \d^2(\bx-\bx')-\sigma^i(\bx')\,\del^i \d^2 (\bx-\bx') \d t 
      \nonumber \\  & &    
           +\frac{1}{2!} \e^{im}(\bx') \e^{jn}(\bx') \nu^m(t) \nu^n(t) 
            \del^i \del^j \d^2 (\bx-\bx')\, \d t^2 
      \nonumber \\  & &     
           +\frac{1}{4!} \g(\bx')\,G_{ab}(\bx')G_{cd}(\bx')
                                   G_{ef}(\bx')G_{gh}(\bx')  \,
            \bar \nu^i_a(t) \bar \nu^j_c(t) \bar \nu^k_e(t) \bar \nu^l_g(t)\, 
      \nonumber \\  & & \;\; 
            \times    
            \z_b(t)\z_d(t)\z_f(t)\z_h(t)   \,
            \del^i \del^j \del^k \del^l \d^2(\bx-\bx')\, \,\d t^4  
      \nonumber \\  & & 
           + \dots 
    \Big \}  \;.
\eea 
After making use of \eref{eq:2nd-corr-nonvanish},
\eref{eq:2nd-bar-corr-nonvanish}, the modified version of 
\eref{eq:4th-corr-nonvanish}, 
\be \langle \z_b(t)\z_d(t)\z_f(t)\z_h(t) \rangle 
   = c_\ng \d_{bd}\,\d_{df}\,\d_{fh} \ , 
\label{eq:4th-corr-nonvanish-II} 
\ee
the notation from~\cite{Blaizot:2002xy},
\be \eta^{ij} (\bx) = \e^{ik}(\bx)\, \e^{jk}(\bx) \ , 
\ee 
and not forgetting that the correlator of four Gaussian noise isn't
zero but can be separated into a sum of products of two two Gaussian noise 
correlators, 
\be  \langle \bar \nu^i_a \bar \nu^j_c \bar \nu^k_e \bar \nu^l_g \rangle_\ga
   = \langle \bar \nu^i_a \bar \nu^j_c \rangle_\ga  
     \langle \bar \nu^k_e \bar \nu^l_g \rangle_\ga      \nonumber \\
    +\langle \bar \nu^i_a \bar \nu^k_e \rangle_\ga  
     \langle \bar \nu^j_c \bar \nu^l_g \rangle_\ga      \nonumber \\
    +\langle \bar \nu^i_a \bar \nu^l_g \rangle_\ga  
     \langle \bar \nu^j_c \bar \nu^k_e \rangle_\ga \ ,      
\label{eq:g4to2g2}
\ee
one gets
\be \frac{\del P(\bx,t)}{\del t} 
   =-\frac{\del \sigma^i(\bx) P(\bx,t)}{\del x^i} 
    +\frac{1}{2!} \frac{\del^2 \eta^{ij} (\bx) P(\bx,t)}{\del x^i \del x^j} 
    -\frac{1}{4!} \frac{\del^4 \Delta^{ijkl}(\bx) P(\bx,t)}
                       {\del x^i \del x^j \del x^k \del x^l} \;, 
\ee 
where $c_\ga = 1/\d t$ has been taken~\footnote{In this case with 
$c_\ga = 1/\d t$ the white noise $\nu^i$ is of the order of 
$\nu^i \sim 1/\sqrt{\d t}$. In terms of the conventional stochastic 
variable that would be $dW^i = \nu^i \dd t \sim \sqrt{\dd t}$ as 
required from stochastic calculus.} and 
\bea \Delta^{ijkl} (\bx) \!\!\!&=&\!\!\! \g(\bx)\,  
     G_{ab}(\bx) G_{cb}(\bx) G_{eb}(\bx) G_{gb}(\bx)         \nonumber \\
  & &\times 
     \Big ( \d_{ac} \d_{eg} \d^{ij} \d^{kl} 
           +\d_{ae} \d_{cg} \d^{ik} \d^{jl}
           +\d_{ag} \d_{ce} \d^{il} \d^{jk}                  
     \Big ) \;.  
\eea 
This equation has all the basic features of \eref{JIMWLK_DM_FP} but some
fine tuning is still required.  This is the subject of the next
subsection.

\subsection{Noise interpretation of the extended JIMWLK equation} 

Back to our present problem, the JIMWLK equation is a functional 
equation of the integrated light-cone gauge field $\a^a$ and the 
evolution is in rapidity $Y$ rather than in time so one has to replace
$t$ by $Y$ everywhere. In analogy to the 
equation for $\dot x^i$ before, we write the stochastic equation 
\be \frac{\del}{\del Y} \a^a(\bu) 
   =  \sigma^a(\bu)+\int_\bz \e^{ab,i}(\bu,\bz) \nu^{b,i}(\bz) 
     +\int_{\bx,\bw,\bz} \rho(\bu,\bx,\bw,\bz) 
      \bar \nu^a(\bx,\bw) \z(\bz) \sqrt{\xi(\bx,\bw)} 
\label{eq:a-stoch-eqn} 
\ee
where borrowing again from the notation
of~\cite{Blaizot:2002xy}~\footnote{Note the slightly different
  notation of the $\a^a$ of~\cite{Blaizot:2002xy} which is equivalent
  to our $\del \a^a /\del Y$.} the  
first stochastic term with the Gaussian noise $\nu^{b,i}$ has the 
tensor coefficient function  
\be \e^{ab,i}(\bu,\bz) = \frac{1}{\sqrt{4\pi^3}} 
                         \frac{(\bu-\bz)^i}{(\bu-\bz)^2}
                         (1-V^\dagger_\bu V_\bz)^{ab}    
\ee
and the product 
\be \eta^{ab} (\bu,\bv) = \int_\bz \e^{ac,i}(\bu,\bz) \e^{bc,i}(\bv,\bz) 
\ee 
from which the first deterministic term $\sigma^a(\bu)$ can be 
expressed as 
\be \sigma^a(\bu) = \frac{1}{2} \int_\bv \frac{\delta}{\delta \a^b(\bv)} 
                    \eta^{ab}(\bu,\bv)   \;.
\ee
The last multi-noise stochastic term on the right hand side of 
\eref{eq:a-stoch-eqn} has a second but different Gaussian noise 
$\bar \nu^a (\bx,\bw)$, the fourth order noise $\z(\bz)$ and a 
new noise $\xi(\bx,\bw)$ which we call a non-diagonal noise for 
reason which will become clear later with a normalized distribution
\be \int \dP_\nd (\xi) = 1  \;. 
\label{eq:nondiagonal-norm} 
\ee 
The coefficient function of this last term is 
\be \rho(\bu,\bx,\bw,\bz) 
   =\sqrt{\frac{\lambda N_c}{(\bx-\bz)^2}}
    \bigg \{ \frac{4!(\bx-\bw)^2 (\nabla^2_\bx\nabla^2_\bw\a^c(\bx)\a^c(\bw)) }
       {3\cdot 2^5 \pi^3 N_g} \bigg \}^{1/4}  {\cal G}(\bu|\bx,\bz) \;. 
\ee 

Here we consider an evolution of the probability density $W_Y[\a]$ of 
the color fields of the CGC by a small increment from $Y' = Y-\d Y$ to $Y$  
\be W_Y[\a] = \int \dD [\a']\, P_{Y|Y'}[\a|\a']\, W_{Y'}[\a'] 
\label{eq:W-cond-prob} 
\ee
during which the fields $\a^a$ make the change 
\be \d \a^a = \a^a-\a^{'a} = \d Y\, \frac{\del}{\del Y} \a^a \;. 
\ee
The conditional probability here is 
\bea P_{Y|Y'}[\a|\a']  
     &=& \int \prod_{a,i} [\DP_\ga(\nu^{a,i})] 
              \prod_b [\dD \bar {\cal P}_\ga(\bar \nu^b)] [\DP_\ng(\z)] 
              [\DP_\nd(\xi)]                                \nonumber \\
     & & \;\; \times \d^{N_g} (\a-\a'-\d Y\, \del \a/\del Y)    \;. 
\label{eq:W-cond-prob-exp} 
\eea 
This is written slightly differently than before because of the 
functional aspect of \eref{eq:W-cond-prob}. The measures of the 
noise $[\DP]$ here are now path integrals for the same reason, and 
$\del \a/\del Y$ is given by \eref{eq:a-stoch-eqn}. Either using
Ito's lemma or expanding the delta function in terms of $\d Y$, 
$W_{Y'}[\a']$ in terms of $\d Y$ as before and substituting in 
\eref{eq:W-cond-prob}, the original JIMWLK equation is 
straightforwardly recovered by following the steps in the previous 
subsection~\cite{Blaizot:2002xy}. The term that is less trivial is 
the four derivative Pomeron splitting term which comes 
essentially from the last term of \eref{eq:a-stoch-eqn} 
raised to the fourth power or the collection of all the apparent
$\d Y^4$ terms in the expansion of the delta function in 
\eref{eq:W-cond-prob-exp}. At this order of the expansion, each 
functional derivative $\d/\d \a^a$ is accompanied by $\z(\bz)$ 
whose correlator ensures that there is only one $z$ coordinate via 
\be \langle \z(\bz)\z(\bz')\z(\bz'')\z(\bz''') \rangle_\ng
   = c_\ng \d^2(\bz-\bz') \d^2(\bz'-\bz'') \d^2(\bz''-\bz''') \ ,  
\ee
where here $c_\ng = -1/\d Y^3$. 
This is the case as seen in \eref{JIMWLK_DM_FP}, the new term has four 
${\cal G}$ functions sharing the same $\bz$ as one endpoint to a
pair of dipoles. This is the result of a one-step evolution from a
single parent dipole. Each of the four ${\cal G}(\bu|\bx,\bz)$'s coming 
from the fourth order expansion of \eref{eq:W-cond-prob-exp} is a 
contribution to the field $\a(\bu)$ from a dipole with endpoints 
$\bx$ and $\bz$. Or equivalently each dipole interacts 
with the target through two gluon exchanges, each of which
is represented by a ${\cal G}(\bu|\bx,\bz)$ therefore the four 
${\cal G}(\bu|\bx,\bz)$'s must come in two pairs with each pair 
sharing the same endpoints. The presence of the second Gaussian 
noise $\bar \nu^a(\bx,\bw)$ correlator 
$\langle \bar \nu^a (\bx_1,\bw_1) \bar \nu^b(\bx_2,\bw_2) 
         \bar \nu^c (\bx_3,\bw_3) \bar \nu^d(\bx_4,\bw_4) \rangle_\ga$,  
which can be rewritten in a similar fashion to \eref{eq:g4to2g2}, 
guarantees that this is the case. These mostly ensure that
one inevitably arrives at \eref{JIMWLK_DM_FP} except the numerator
of the BFKL kernel, which is $(\bx-\by)^2$. Naively one is required
to put the quartic root of this expression in the stochastic equation
for $\a^a$ in \eref{eq:a-stoch-eqn} which is impossible since only
one of $\bx$ and $\by$ can appear in the expression. The other comes
into \eref{JIMWLK_DM_FP} via the $\d Y$ expansion and the $\bar \nu$ 
correlator discussed above. Thus to ensure that the appearance of the
correct BFKL kernel, we gave the factor ${\{(\bx-\bw)^2\}}^{1/4}$ 
to \eref{eq:a-stoch-eqn}, introduced the very unconventional 
non-diagonal noise $\sqrt{\xi(\bx,\bw)}$ and required the latter to 
satisfy the correlation  
\be  \langle \xi(\bx,\bw) \xi(\by,\bw') \rangle_\nd
   = \d^2(\bx-\bw') \d^2(\by-\bw) \;. 
\label{eq:nondiagonal-corr}
\ee
One should note the unusual pairing of the arguments of the 
$\d$-functions. This is the reason for its nomenclature. 
Here two sets of coordinate pair have already been identified as 
$\bx, \bw$ and $\by, \bw'$ thanks to the $\bar \nu^a$ correlators
so the square root has been removed. This correlation is somewhat 
counterintuitive because one encounters much more frequently the 
normal Gaussian noise, whose correlator is  
\be  \langle \xi(\bx,\bw) \xi(\by,\bw') \rangle_\ga 
   = \d^2(\bx-\by) \d^2(\bw-\bw')   \;.
\ee
As it is the non-diagonal correlator requires that a squared quantity 
to be zero  
\be  \langle \xi(\bx,\bw)^2 \rangle_\nd = 0   \;. 
\label{eq:square-zero}
\ee
This demands a lot on the distribution itself. The only constraint 
on $\xi(\bx,\bw)$ here is \eref{eq:nondiagonal-norm} and 
\eref{eq:nondiagonal-corr}. Note that although $\sqrt{\xi(\bx,\bw)}$
appears in the stochastic equation \eref{eq:a-stoch-eqn}, one does
not encounter correlators such as 
\[ \langle \sqrt{\xi(\bx,\bw)} \rangle_\nd \;\;,\;\; 
   \langle \sqrt{\xi(\bx,\bw)\xi(\bx',\bw')} \rangle_\nd 
\]
etc because of the presence of $\bar \nu^a(\bx,\bw)$ and $\z(\bx)$.
Such correlators are not only unpleasant but also not very meaningful.
We are unable to find such an unusual correlator in the mathematics
literature but if the stochastic interpretation of \eref{JIMWLK_DM_FP} is
to be preserved, one would have to require such a correlator to exist.
Note that the vanishing of the average over a squared quantity similar
to \eref{eq:square-zero} is also required of the fourth order noise
$\z$ for which one knows that a distribution does exist. Providing one
accepts that a distribution for $\xi$ might probably exist then with
all the ingredients in place \eref{JIMWLK_DM_FP} can be reproduced from
the multi-noise stochastic equation \eref{eq:a-stoch-eqn} by going
through the steps as shown in this section. So once again although the
extended JIMWLK equation for a dilute color dipole medium appears to
be more complicated with a fourth order functional derivative term so
that the Langevin description is no longer an option, nevertheless one
can preserve the stochastic interpretation through the introduction of
higher order noise. The original goal for the Langevin description is
to facilitate the computation of the evolution of high energy
collisions with CGC through the JIMWLK equation. In a dilute medium
the new term is as important as any terms in the original JIMWLK
equation, however computationwise, it is not clear that the new
stochastic equation here necessary for the new development would
fulfill its original purpose since it might be difficult, if not
impossible, to implement both the fourth order noise and the
non-diagonal noise on a computer. Again provided a distribution
exists, the last would likely to pose the most difficulties.  Should
this distribution turn out eventually to be not possible then the
stochastic description is really lost.  We will leave that for future
consideration and discussion.


\section*{Acknowledgements}

We thank Edmond Iancu for many enlightening discussions on this
general problem and for explaining to us the ideas and details of his
work with D. Triantafyllopoulos. We thank the latter for an elegant
note showing the equivalence between their approach and ours. A.~Sh.
acknowledges financial support by the Deutsche Forschungsgemeinschaft
under contract Sh 92/1-1.

%
%
  
%
%
\end{document}